\documentclass[preprint]{elsarticle}

\usepackage[cmex10]{amsmath}
\usepackage{amssymb}
\usepackage{amsthm }
\usepackage{nicefrac}

\newtheorem{theorem}{Theorem}

\theoremstyle{theorem}

\theoremstyle{definition}

\theoremstyle{plain}

\theoremstyle{plain}

\usepackage{hyperref}
\usepackage[bottom]{footmisc}
\usepackage{url}
\usepackage{bm}
\usepackage{xcolor}
\hypersetup{colorlinks=true}
\usepackage{tabularx}

\usepackage{algorithm}
\usepackage{multirow}
\usepackage{algcompatible}
\usepackage{adjustbox}

\usepackage{epsfig,epsf,epstopdf,graphicx}
\usepackage[caption=false,font=footnotesize]{subfig}

\usepackage{fixltx2e}
\usepackage{afterpage}
\usepackage{float}
\usepackage{stfloats}

\newcommand{\cb}{{\bf{c}}}

\newcommand{\Fb}{\mathbf{F}}

\newcommand{\Ub}{\mathbf{U}}

\newcommand{\vb}{\mathbf{v}}

\newcommand{\Xb}{\mathbf{X}}

\newcommand{\thetab}{\mathbf{\theta}}
\newcommand{\omegab}{\mathbf{\omega}}
\newcommand{\etab}{\mathbf{\eta}}
\newcommand{\varphib}{\mathbf{\varphi}}
\newcommand{\me}{{\mathcal{E}}}

\newcommand{\xb}{\mathbf{x}}
\newcommand{\ub}{\mathbf{u}}
\newcommand{\wb}{\mathbf{w}}
\newcommand{\pb}{\mathbf{p}}
\newcommand{\qb}{\mathbf{q}}

\newcommand{\tb}{\mathbf{t}}
\newcommand{\eb}{\mathbf{e}}
\newcommand{\fb}{\mathbf{f}}
\newcommand{\gb}{\mathbf{g}}
\newcommand{\kb}{\mathbf{k}}
\newcommand{\rb}{\mathbf{r}}
\newcommand{\hb}{\mathbf{h}}

\newcommand{\bb}{\mathbf{b}}
\newcommand{\db}{\mathbf{d}}
\newcommand{\zb}{\mathbf{z}}
\newcommand{\sbb}{\mathbf{s}}

\newcommand\Tstrut{\rule{0pt}{2.6ex}}         

\journal{Journal of DSP Signal Processing}

\begin{document}
\begin{frontmatter}
\title{Second-Order Nonlinearity Estimated and Compensated Diffusion LMS Algorithm: Theoretical Upper Bound, Cramer-Rao Lower bound, and Convergence Analysis}

\author[mymainaddress]{Hadi~Zayyani\corref{mycorrespondingauthor}}
\ead{zayyani@qut.ac.ir}
\cortext[mycorrespondingauthor]{Corresponding author}

\author[mysecondaryaddress]{Mehdi~Korki}
\ead{mkorki@swin.edu.au}



\address[mymainaddress]{Qom University of Technology (QUT), Qom, Iran}
\address[mysecondaryaddress]{Swinburne University of Technology, Melbourne, Australia}

\begin{abstract}
In this paper, an algorithm for estimation and compensation of second-order nonlinearity in wireless sensor setwork (WSN) in distributed estimation framework is proposed. First, the effect of second-order nonlinearity on the performance of Diffusion Least Mean Square (DLMS) algorithm is investigated and an upper bound for $l^2$-norm of the error due to nonlinearity is derived mathematically. Second, mean convergence analysis of the DLMS algorithm in presence of second-order nonlinearity is derived. Third, a distributed algorithm is suggested which consists of extra nonlinearity estimation and compensation units. Moreover, considering the second-order nonlinearity, the Cramer-Rao bound (CRB) for estimating both the unknown vector and nonlinearity coefficient vector is calculated, in which the Fisher information matrix is obtained in a closed-form formula. Simulation results demonstrate the effectiveness of the proposed algorithm in improving the performance of distributed estimation in the presence of nonlinear sensors in a WSN.
\end{abstract}


\end{frontmatter}

\section{Introduction}
\label{sec:Intro}
 The problem of distributed estimation of an unknown vector from linear measurements is a well-known subject in signal processing community which has numerous applications in wireless sensor network (WSN), channel estimation, spectrum estimation, massive MIMO communication, and target tracking problems \cite{Sayed14}, \cite{Pasha16}. The distributed estimation algorithms benefit from the inter-collaboration of sensor nodes. The cooperation strategies for distributed estimation are incremental, consensus, and diffusion approaches \cite{Sayed14}. Among them, the diffusion strategy is more versatile due to its simplicity, scalability, and low storage demands.

Many distributed diffusion algorithms are proposed in the literature, e.g., diffusion LMS \cite{LopS08}, \cite{CatS10}, diffusion LMP \cite{Wen13}, \cite{Lu18}, diffusion Affine Projection Algorithm (APA) \cite{Shams19}, \cite{Shiri18}, diffusion CMPN \cite{Kork19}, and diffusion correntropy \cite{Ma16}-\cite{Gogi20},  to name a few. Among diffusion algorithms, the Diffusion Least Mean Square (DLMS) algorithm is the basic algorithm which uses mean square error (MSE) as its cost function. There are also numerous variants of the DLMS algorithm, which aim to either reduce the communication load \cite{Arab14}, \cite{Zayy22CommRed}, make the algorithm robust against impulsive noise \cite{Zayy20}, \cite{ZayyJ21}, make the algorithm secure with respect to adversaries \cite{Chang20}, \cite{Zayy22CSSP}, \cite{Zayy23}, or in sparse setting \cite{Moda20}. Unfortunately, the performance of the aforementioned algorithms in a WSN deteriorates when the sensors have some nonlinearity effect due e.g. to their power amplifiers. This is because they are designed for the linear measurement model. The main objective of this paper is to make the DLMS algorithm robust against nonlinearities.

In the literature of distributed estimation, there are some works that consider a nonlinear model for the measurements \cite{Kar12}-\cite{Meng21}. In the pioneering work of \cite{Kar12}, two distributed algorithms are suggested for estimation in a nonlinear observation model. Moreover, a diffusion based kernel least mean squares (KLMS) is presented in a nonlinear measurement setup \cite{Chouv16}. In addition, a distributed estimation algorithm with nonlinear sensors with one bit measurements are proposed in \cite{Zayy17}. Besides, \cite{Chen20} suggests two algorithms for estimating the parameters of nonlinear Hammerstein systems with missing data. A distributed nonlinear parameter estimation algorithm is further developed in unbalanced multi-agent networks \cite{Meng20}. Nonlinear model is partially used in \cite{Chan19}, in which a method for distributed solution of robust estimation problems is proposed with equality constraints based on the augmented Lagrangian method. \cite{Meng21} discusses both linear and nonlinear models for secure distributed estimation in the presence of attackers in the network.

In this paper, we deal with second-order non-linear model for sensors. It allows to model a linear system which shows some small degree of nonlinearities. The challenges of the second-order nonlinear model is in adaptiveness in which the nonlinear coefficient may change over time. In the proposed solution, the adaptiveness are taken into account. Thanks to the second-order nonlinear model, we can investigate the nonlinearity effect on the performance of a DLMS algorithm. Hence, an upper bound for the error is calculated in the paper. Also, an improved version of DLMS algorithm is suggested which incorporates nonlinearity estimation and compensation units. Further, the Cramer-Rao bound (CRB) is calculated for the distributed estimation problem in the presence of second-order nonlinearity. Simulation results show the benefit of the proposed method especially when there are small nonlinear coefficients.


\section{System model and problem formulation}
\label{sec:ProblemForm}

Consider a WSN with $N$ sensors (nodes) collecting a scalar measurement $d_{k,i}$,  where $1\leq k \leq N$ is the node index and $1\le i\le I$ is the time instant. Each sensor contains its own $L\times 1$ input regression vector $\ub_{k,i}$. The model of measurements is linear, i.e., $d_{k,i}={\ub^T_{k,i}}\omegab_o+v_{k,i}$, with the unknown $L\times 1$ vector $\omegab_o$, where $v_{k,i}$ denotes the measurement noise. It is assumed that the sensor equipment has a power amplifier with a second-order nonlinear model. So, if the nonlinear function is $f(\cdot)$ then nonlinear measurements are
\begin{equation}
\label{eq: dtilde}
\tilde{d}_{k,i}=f(d_{k,i})=d_{k,i}+b_k d^2_{k,i}+\theta_{k,i},
\end{equation}
where $b_k$ is the second-order nonlinearity coefficient of $k$th sensor and $\theta_{k,i}$ is the measurement noise which is assumed to be zero-mean Gaussian with variance $\sigma^2_{\theta,k}$. The constant term in the above model (\ref{eq: dtilde}) is omitted since the system is an approximately linear system with a second-order nonlinear term.

The main objective of the distributed estimation problem in the WSN is to estimate the unknown vector $\omegab_o$ using nonlinear measurements $\tilde{d}_{k,i}$ and regression vectors $\ub_{k,i}$ of sensors. The other objective is to estimate the nonlinearity of the sensors in the network.

\section{The DLMS Algorithm in the presence of nonlinearities}
\label{sec: DLMS}
Diffusion algorithms are usually suitable solutions for distributed estimation problems, of which DLMS is the most basic. The two steps of the DLMS algorithm is the adaptation and combination steps. It can be implemented in two ways: Adapt Then Combine (ATC) and Combine Then Adapt (CTA). The ATC version of DLMS is as follows \cite{Sayed14}:
\begin{equation}
\label{eq: DLMS}
\Bigg\{\begin{array}{ll}
\tilde{\varphib}_{k,i}=\omegab_{k,i-1}+\mu_k\sum_{l\in \cal N_\mathrm{k}}c_{lk}\ub_{l,i}(\tilde{d}_{l,i}-\ub^T_{l,i}\omegab_{k,i-1}),\\
\tilde{\omegab}_{k,i}=\sum_{l\in \cal N_\mathrm{k}}a_{lk}\tilde{\tilde{\varphib}}_{l,i},
                  \end{array}
\end{equation}
where $\cal N_\mathrm{k}$, $\tilde{\varphib}_{k,i}$, and $\tilde{\tilde{\varphib}}_{l,i}=f(\tilde{\varphib}_{l,i})$ denote the neighborhood set of the $k$'th sensor, the intermediate estimation of $k$'th sensor in the presence of nonlinearity at time index $i$, and the received intermediate estimation of sensor $l$, respectively. Further, $a_{lk}$ and $c_{lk}$ are the combination coefficients from node $l$ to node $k$ in the adaptation and combination steps, respectively. The local cost function of node $k$ in the DLMS algorithm in presence of nonlinearity is defined as
\begin{equation}
\label{eq: costf}
\tilde{J}_k(\omega)=\sum_{l\in \cal N_\mathrm{k}}c_{lk}\mathrm{E}\{||\tilde{d}_{l,i}-\ub^T_{l,i}\omegab||^2_2\}.
\end{equation}
where expectation operator $\mathrm{E}\{\}$ can be neglected and the point estimation can be replaced for expectation.

\section{The DLMS Algorithm in the presence of nonlinearities}
\label{sec: DLMS}
Diffusion algorithms are usually suitable solutions for distributed estimation problems, of which DLMS is the most basic. The two steps of the DLMS algorithm is the adaptation and combination steps. It can be implemented in two ways: Adapt Then Combine (ATC) and Combine Then Adapt (CTA). The ATC version of DLMS is as follows \cite{Sayed14}:
\begin{equation}
\label{eq: DLMS}
\Bigg\{\begin{array}{ll}
\tilde{\varphib}_{k,i}=\omegab_{k,i-1}+\mu_k\sum_{l\in \cal N_\mathrm{k}}c_{lk}\ub_{l,i}(\tilde{d}_{l,i}-\ub^T_{l,i}\omegab_{k,i-1}),\\
\tilde{\omegab}_{k,i}=\sum_{l\in \cal N_\mathrm{k}}a_{lk}\tilde{\tilde{\varphib}}_{l,i},
                  \end{array}
\end{equation}
where $\cal N_\mathrm{k}$, $\tilde{\varphib}_{k,i}$, and $\tilde{\tilde{\varphib}}_{l,i}=f(\tilde{\varphib}_{l,i})$ denote the neighborhood set of the $k$'th sensor, the intermediate estimation of $k$'th sensor in the presence of nonlinearity at time index $i$, and the received intermediate estimation of sensor $l$, respectively. Further, $a_{lk}$ and $c_{lk}$ are the combination coefficients from node $l$ to node $k$ in the adaptation and combination steps, respectively. The local cost function of node $k$ in the DLMS algorithm in presence of nonlinearity is defined as
\begin{equation}
\label{eq: costf}
\tilde{J}_k(\omega)=\sum_{l\in \cal N_\mathrm{k}}c_{lk}\mathrm{E}\{||\tilde{d}_{l,i}-\ub^T_{l,i}\omegab||^2_2\}.
\end{equation}
where expectation operator $\mathrm{E}\{\}$ can be neglected and the point estimation can be replaced for expectation.

\section{The Upper bound for the error term due to nonlinearity}
\label{sec: upper}
In this section, we investigate the effect of second-order nonlinearity on the performance of the DLMS algorithm. We derive an upper bound for the $l^2$-norm of the error term, which is the difference between the estimated vector after combination step in the presence of nonlinearity and without the nonlinearity. To that end, we write the formula of the intermediate estimation in (\ref{eq: DLMS}) in the following form
\begin{displaymath}
\tilde{\varphib}_{k,i}=\omegab_{k,i-1}-\mu_k\nabla_{\omegab}\tilde{J}(\omegab_{k,i-1})=
\end{displaymath}
\begin{displaymath}
\omegab_{k,i-1}+\mu_k\sum_{l\in \cal N_\mathrm{k}}c_{lk}\ub_{l,i}(\tilde{d}_{l,i}-\ub^T_{l,i}\omegab_{k,i-1})=\\
\end{displaymath}
\begin{equation}
\varphib_{k,i}+\mu_k \sum_{l\in \cal N_\mathrm{k}}c_{lk}(b_ld^2_{l,i})\ub_{l,i}=\varphib_{k,i}+\Delta\varphib_{k,i},
\end{equation}
where the noise term $\theta_l$ is neglected in comparison to nonlinear term $b_ld^2_{l,i}$ and $\Delta\varphib_{k,i}\triangleq \mu_k \sum_{l\in \cal N_\mathrm{k}}c_{lk}(b_l\tilde{d}^2_{l,i})\ub_{l,i}$ is the error vector of the intermediate estimation. In fact, we assume that the noise level is much lower than the nonlinearity term. Then, the intermediate estimations of $\tilde{\varphib}_{l,i}$ is received by the node $k$ as $\tilde{\tilde{\varphib}}_{l,i}=f(\tilde{\varphib}_{k=l,i})+\eta_{l,i}$. So, we have
\begin{displaymath}
\tilde{\tilde{\varphib}}_{l,i}=\tilde{\varphib}_{l,i}+b_l\tilde{\varphib}^2_{l,i}+\eta_{l,i}=
\end{displaymath}
\begin{equation}
\varphib_{l,i}+\Delta\varphib_{l,i}+b_l(\varphib_{l,i}+\Delta\varphib_{l,i})^2+\eta_{l,i},
\end{equation}
where $\xb^2\triangleq \xb\odot\xb$ is the element-wise square of a vector in which $\odot$ is the hadamard operator. Then, the output of combination unit will be
\begin{displaymath}
\tilde{\omega}_{k,i}=\sum_{l\in \cal N_\mathrm{k}}a_{lk}\tilde{\tilde{\varphib}}_{l,i}=
\end{displaymath}
\begin{equation}
\omega_{k,i}+\sum_{l\in \cal N_\mathrm{k}}a_{lk}\Big[\Delta\varphib_{l,i}+b_l(\varphib_{l,i}+\Delta\varphib_{l,i})^2+\eta_{l,i}\Big],
\end{equation}
where $\omega_{k,i}=\sum_{l\in \cal N_\mathrm{k}}a_{lk}\varphib_{l,i}$ is the new true estimation without nonlinearity. To upper bound the $l^2$-norm of the error vector of $\tilde{\omega}_{k,i}-\omega_{k,i}$, we write
\begin{displaymath}
\me=||\tilde{\omega}_{k,i}-\omega_{k,i}||^2_2=
\end{displaymath}
\begin{displaymath}
\sum_{l_1\in \cal N_\mathrm{k}}\sum_{l_2\in \cal N_\mathrm{k}}a_{l_1,k}a_{l_2,k}\Big[\Delta\varphib_{l_1,i}+b_{l_1}(\varphib_{l_1,i}+\Delta\varphib_{l_1,i})^2\Big]^T
\end{displaymath}
\begin{equation}
\label{eq: Eterm}
\quad\quad\quad\Big[\Delta\varphib_{l_2,i}+b_{l_2}(\varphib_{l_2,i}+\Delta\varphib_{l_2,i})^2\Big].
\end{equation}

The upper bound for the error term is expressed in the following theorem.

	\begin{theorem}
		\label{theorem1}
		The error term $\me=||\tilde{\omega}_{k,i}-\omega_{k,i}||^2_2$ is upper bounded by
	\end{theorem}
	\begin{equation}
		\me=||\tilde{\omega}_{k,i}-\omega_{k,i}||^2_2\le C_{1,\mathrm{max}}\sum_{l_1\in \cal N_\mathrm{k}}\sum_{l_2\in \cal N_\mathrm{k}}a_{l_1,k}a_{l_2,k},
	\end{equation}
	where
    \begin{displaymath}
    C_{1,\mathrm{max}}=2b^{2.5}_{l,\mathrm{max}}L\mu^{1.5}||\omegab_o||^3+
    \end{displaymath}
    \begin{equation}
    \label{eq: C1}
    \mu^{1.5} b^{3.5}_{l,\mathrm{max}}||\omegab_o||^3(L\sqrt{\mu b_{l,\mathrm{max}}}||\omegab_o||+4\sqrt{L})^2,
    \end{equation}
    where $b_{l,\mathrm{max}}$ is the upper bound of $|b_l|$.

	\begin{proof}
Neglecting $i$ and $k$ in (\ref{eq: Eterm}) for simplicity, and assuming the nonlinear coefficients $b_l$ are small, we can neglect the second order error terms, i.e., $\Delta\varphib^T_{l_1}\Delta\varphib_{l_2}$. Then, we have the following approximation
\begin{displaymath}
\me\approx\sum_{l_1}\sum_{l_2}a_{l_1}a_{l_2}\Big\{b_{l_2}\Delta\varphib^T_{l_1}\varphib^2_{l_2}+b_{l_1}(\varphib^2_{l_1})^T\Delta\varphib_{l_2}+b_{l_1}b_{l_2}\Big[
\end{displaymath}
\begin{displaymath}
(\varphib^2_{l_1})^T\varphib^2_{l_2}+2\Big(\varphib^T_{l_1}\odot\Delta\varphib^T_{l_1}\Big)\varphib^2_{l_2}+2(\varphib^2_{l_1})^T\Big(\varphib_{l_2}\odot\Delta\varphib_{l_2}\Big)\Big]\Big\}
\end{displaymath}
\begin{equation}
=\sum_{l_1}\sum_{l_2}a_{l_1}a_{l_2}\Psi_{l_1,l_2}.
\end{equation}

Using the triangular inequality, we have
\begin{equation} 
\me\approx \sum_{l_1}\sum_{l_2}a_{l_1}a_{l_2}\Psi_{l_1,l_2}\le \sum_{l_1}\sum_{l_2}a_{l_1}a_{l_2}|\Psi_{l_1,l_2}|,
\end{equation}
and we achieve
\begin{displaymath}
|\Psi_{l_1,l_2}|\le|b_{l_2}||\Delta\varphib^T_{l_1}\varphib^2_{l_2}|+|b_{l_1}||(\varphib^2_{l_1})^T\Delta\varphib_{l_2}|+
\end{displaymath}
\begin{equation}
b_{l_1}b_{l_2}\Big[|(\varphib^2_{l_1})^T\varphib^2_{l_2}|+2|(\varphib^T_{l_1}\odot\Delta\varphib^T_{l_1})\varphib^2_{l_2}|+2|(\varphib^2_{l_1})^T(\varphib_{l_2}\odot\Delta\varphib_{l_2})|\Big].
\end{equation}

Now, we assume $|\Delta\varphib_j|^2<M_j$, where $M_j$ is the upper bound for square error of elements. Also, we assume $||\Delta\varphib||^2_2<M_{\Delta\phi}$, where $M_{\Delta\phi}$ is the upper bound of the $l^2$-norm. We also know that $M_j$ and $M_{\Delta\phi}$ have linear relationship, i.e., $M_{\Delta\phi}=LM_j$. Thus, we only derive the upper bound $M_j$. By neglecting the noise term and assuming $\mu_k=\mu$, we can write
\begin{equation}
|\Delta\varphib_j|=|\mu\sum_{l\in \cal N_\mathrm{k}}c_{lk}(b_l{d}^2_{l,i})u_{l,i,j}|\le\mu\sum_{l\in \cal N_\mathrm{k}}c_{lk}|b_l||{d}^2_{l,i}||u_{l,i,j}|.
\end{equation}

Assuming $||\ub_{l,i}||=1$, without loss of generality, we have $|u_{l,i,j}|\le 1$. Also, using Cauchy-Schuartz inequality, we have $|d_{l,i}|^2=|\ub^T_{l,i}\omegab_o|\le||\omegab_o||^2$. Then, we have
\begin{equation}
|\Delta\varphib_j|\le\mu\sum_{l\in \cal N_\mathrm{k}}c_{lk}b_{l,{\mathrm{max}}}||\omegab_o||^2=\mu b_{l,{\mathrm{max}}}||\omegab_o||^2=M_j=M.
\end{equation}
Further, we have $M_{\Delta\phi}=\mu L b_{l,{\mathrm{max}}}||\omegab_o||^2$. After finding the upper bounds $M_j$ and $M_{\Delta\phi}$, we derive the upper bound for $|\Psi_{l_1,l_2}|$ by applying Cauchy-Schartz inequality. Hence, we have
\begin{displaymath}
|\Psi_{l_1,l_2}|\le|b_{l_2}||\Delta\varphib_{l_1}||.||\varphib^2_{l_2}||+|b_{l_1}||\Delta\varphib_{l_2}||.||\varphib^2_{l_1}||+b_{l_1}b_{l_2}\Big[
\end{displaymath}
\begin{equation}
||\varphib^2_{l_1}||.||\varphib^2_{l_2}||+2||\varphib^T_{l_1}\odot\Delta\varphib^T_{l_1}||.||\varphib^2_{l_2}||+2||\varphib^T_{l_2}\odot\Delta\varphib^T_{l_2}||.||\varphib^2_{l_1}||\Big].
\end{equation}
Then, considering the upper bounds, we have
\begin{displaymath}
|\Psi_{l_1,l_2}|\le b_{l,\mathrm{max}}\sqrt{M_{\Delta\phi}}(||\varphib^2_{l_1}||+||\varphib^2_{l_2}||)+b^2_{l,\mathrm{max}}\Big[
\end{displaymath}
\begin{equation}
\label{eq: psi12}
||\varphib^2_{l_1}||||\varphib^2_{l_2}||+2||\varphib^2_{l_2}||B_1+2||\varphib^2_{l_1}||B_2\Big],
\end{equation}
where $B_1$ and $B_2$ are the upper bounds of $||\varphib^T_{l_1}\odot\Delta\varphib^T_{l_1}||$ and $||\varphib^T_{l_2}\odot\Delta\varphib^T_{l_2}||$, respectively. We have $||\varphib^T_{l_1}\odot\Delta\varphib^T_{l_1}||^2=\sum_{j}(\phi_{l_1,j}\Delta\phi_{l_1,j})^2\le M_j||\varphib||^2=M_j=B^2_1$, where without loss of generality, we assume that the intermediate estimations are normalized to unity. Similarly, we have $B_2=\sqrt{M_j}$. Besides, we have $||\varphib^2_{l_1}||=\sqrt{\sum_{j=1}^L \phi^4_{l_1,j}}\le M_j\sqrt{L}$. Hence, following (\ref{eq: psi12}) and simplifying the terms, we have
\begin{displaymath}
|\Psi_{l_1,l_2}|\le 2b_{l,\mathrm{max}}L(M_j)^{1.5}+b^2_{l,\mathrm{max}}L(M_j)^{1.5}(L\sqrt{M_j}+4\sqrt{L}).
\end{displaymath}
\begin{equation}
\label{eq: boundf}
=C_{1,\mathrm{max}}
\end{equation}
Then, substituting $M_j=\mu b_{l,{\mathrm{max}}}||\omegab_o||^2$ in (\ref{eq: boundf}), with some simplifications, the proof is achieved.
	\end{proof}


\section{Mean convergence analysis in presence of nonlinearity}
\label{sec: MC}
In this section, the mean convergence analysis of DLMS in presence of second-order nonlinearity is performed. In the first case, we assume the nonlinearity both in measurements and links. In the second case, we consider the nonlinearity just in measurements.

\subsection{Genral-case:nonlinearity both in measurements and links}
From the combination step of the DLMS algorithm in presence of second-order nonlinearity, we have
\begin{displaymath}
\omegab_{k,i}=\sum_{l\in \cal N_\mathrm{k}}a_{lk}\Big[\tilde{\varphib_{l,i}}+b_l\tilde{\varphib^2_{l,i}}+\etab_{l,i}\Big]=
\end{displaymath}
\begin{displaymath}
\sum_{l\in \cal N_\mathrm{k}}a_{lk}\Big[\tilde{\varphib_{l,i}}+b_l(\varphib_{l,i}+\Delta\varphib_{l,i})+\etab_{l,i}\Big]=
\end{displaymath}
\begin{equation}
\sum_{l\in \cal N_\mathrm{k}}a_{lk}\varphib_{l,i}+\sum_{l\in \cal N_\mathrm{k}}a_{lk}\Big[\tilde{\varphib_{l,i}}+b_l(\varphib_{l,i}+\Delta\varphib_{l,i})+\etab_{l,i}\Big],
\end{equation}
where $\Delta\varphib_{l,i}$ and $\varphib_{l,i}$ are given by
\begin{equation}
\Delta\varphib_{l,i}\triangleq \mu \sum_{l^{'}\in \cal N_\mathrm{l}}c_{l^{'}k}(b_l^{'}\tilde{d}^2_{l^{'},i})\ub_{l^{'},i},
\end{equation}
and
\begin{equation}
\varphib_{l,i}=\omegab_{l,i-1}+\mu\pb_{l,i},
\end{equation}
where $\pb_{l,i}=\sum_{l^{'}\in \cal N_\mathrm{l}}c_{l^{'}k}e_{l^{'},i}\ub_{l^{'},i}$ in which we have $e_{l^{'},i}=d_{l^{'},i}-\ub^{l^{'},i}\omegab_{l^{'},i-1}$. So, if we define $\tilde{\omegab}_{l,i}=\omegab_{l,i}-\omegab^{o}$ and $\tilde{\tilde{\omegab}}_{l,i}=\mathrm{E}\{\tilde{\omegab}_{l,i}\}$, we have
\begin{equation}
\label{eq: mco}
\tilde{\tilde{\omegab}}_{l,i}=\sum_{l\in \cal N_\mathrm{l}}a_{lk}\tilde{\tilde{\omegab}}_{l,i-1}+\sum_{l\in \cal N_\mathrm{l}}a_{lk}\Big[\mu\mathrm{E}\{\pb_{l,i}\}+\mathrm{E}\{\Delta\varphib_{l,i}\}+b_l\mathrm{E}\{(\varphib_{l,i}+\Delta\varphib_{l,i})^2\}\Big].
\end{equation}
Then, if we define $\fb_{l,i}\triangleq\mathrm{E}\{\pb_{l,i}\}$, $\gb_{l,i}\triangleq\mathrm{E}\{\Delta\varphib_{l,i}\}$, and $\kb_{l,i}\triangleq\mathrm{E}\{(\varphib_{l,i}+\Delta\varphib_{l,i})^2\}$, then (\ref{eq: mco}) can be written in the following form
\begin{equation}
\label{eq: com}
\tilde{\tilde{\omegab}}_{l,i}=\sum_{l\in \cal N_\mathrm{l}}a_{lk}\tilde{\tilde{\omegab}}_{l,i-1}+\sum_{l\in \cal N_\mathrm{l}}a_{lk}\Big[\mu\fb_{l,i}+\gb_{l,i}+b_l\kb_{l,i}\Big].
\end{equation}
In the appendix 1, the $\fb_{l,i}$, $\gb_{l,i}$, and $kb_{l,i}$ are calculated as
\begin{equation}
\label{eq: fli}
\fb_{l,i}=-\sigma^2_u\sum_{l^{'}\in \cal N_\mathrm{l}}c_{l^{'},l}\tilde{\tilde{\omegab}}_{l^{'},i-1},
\end{equation}
,
\begin{equation}
\label{eq: gli}
\gb_{l,i}=\mathbf{0},
\end{equation}
and
\begin{equation}
\label{eq: kli}
\kb_{l,i}=\mathrm{E}\{\varphib^2_{l,i}\}+\mathrm{E}\{\Delta\varphib^2_{l,i}\}=\hb_{l,i}+\rb_{l,i},
\end{equation}
where $\hb_{l,i}\triangleq\mathrm{E}\{\varphib^2_{l,i}\}$ and $\rb_{l,i}=\mathrm{E}\{\Delta\varphib^2_{l,i}\}$.
In appendix 2, the $\hb_{l,i}$ and $\rb_{l,i}$ are computed in Appendix 2.

Now, putting all together, (\ref{eq: com}) can be written as
\begin{displaymath}
\tilde{\tilde{\omegab}}_{l,i}=\sum_{l\in \cal N_\mathrm{l}}a_{lk}\tilde{\tilde{\omegab}}_{l,i-1}-\mu\sigma^2_u\sum_{l\in \cal N_\mathrm{l}}a_{lk}\sum_{l^{'}\in \cal N_\mathrm{l}}c_{l^{'},l}\tilde{\tilde{\omegab}}_{l^{'},i-1}+\sum_{l\in \cal N_\mathrm{l}}a_{lk}b_l(\hb_{l,i}+\rb_{l,i})=
\end{displaymath}
\begin{equation}
\label{eq: recf}
\sum_{l\in \cal N_\mathrm{l}}\gamma_{lk}\tilde{\tilde{\omegab}}_{l,i-1}+\gb_{k,i},
\end{equation}
where $\gamma_{l^{'}k}=a_{l^{'}k}-\mu\sigma^2_u\sum_{l\in \cal N_\mathrm{l^{'}}}a_{lk}c_{l^{'},l}$ and we have
\begin{equation}
\gb_{k,i}=\sum_{l\in \cal N_\mathrm{l}}a_{lk}b_l(\hb_{l,i}+\rb_{l,i}).
\end{equation}

\subsection{Special-case:nonlinearity only in measurements}
In this part, the mean convergence analysis is performed when the only nonlinearity is in measurements. In this case, we can write
\begin{displaymath}
\omegab_{k,i}=\sum_{l\in \cal N_\mathrm{l}}a_{lk}\tilde{\varphib}_{l,i}=\sum_{l\in \cal N_\mathrm{l}}a_{lk}(\varphib_{l,i}+\Delta\varphib_{l,i})=
\end{displaymath}
\begin{equation}
\label{eq: 222}
\sum_{l\in \cal N_\mathrm{l}}a_{lk}(\omegab_{l,i-1}+\mu\pb_{l,i}+\Delta\varphib_{l,i}).
\end{equation}
Then, (\ref{eq: 222}) can be written as
\begin{equation}
\label{eq: 333}
\omegab_{k,i}-\omegab^{o}=\sum_{l\in \cal N_\mathrm{l}}a_{lk}(\omegab_{l,i-1}-\omegab^{o})+\sum_{l\in \cal N_\mathrm{l}}a_{lk}(\mu\pb_{l,i}+\Delta\varphib_{l,i}).
\end{equation}
By taking expectation from both sides of (\ref{eq: 333}), we have
\begin{equation}
\label{eq: finrec}
\tilde{\tilde{\omegab}}_{k,i}=\sum_{l\in \cal N_\mathrm{l}}a_{lk}\tilde{\tilde{\omegab}}_{l,i-1}-\mu\sigma^2\sum_{l\in \cal N_\mathrm{l}}a_{lk}\sum_{l^{'}\in \cal N_\mathrm{l}}c_{l^{'},l}\tilde{\tilde{\omegab}}_{l^{'},i-1}.
\end{equation}
Now, (\ref{eq: finrec}) can be written as a recursion without bias term, as follows:
\begin{equation}
\tilde{\tilde{\omegab}}_{k,i}=\sum_{l\in \mathrm{Graph}}\tilde{a}_{l,k}\tilde{\tilde{\omegab}}_{l,i-1},
\end{equation}
where $\tilde{a}_{l,k}=a_{lk}-\mu\sigma^2_u\sum_{l^{'}\in \cal N_\mathrm{l}}c_{l^{'},l}$.


\section{The Proposed Algorithm}
\label{sec: prop}
DLMS algorithm performance is degraded by nonlinearity. To improve the performance of the DLMS algorithm in the presence of nonlinearity, we propose to estimate the nonlinear coefficients and then compensate their effects. This process is adaptive and online as it shows its benefits in comparison to a pre-calibration process. We call the proposed algorithm second-order nonlinearity estimated and compensated DLMS (SONEC-DLMS) algorithm. It consists of five different steps. The details of steps are:
\begin{enumerate}
\item Nonlinear coefficient estimation: If we define the coefficient vector of node $k$ as $\bb_k=[b_l]_{l\in \cal N_\mathrm{k}}$, we estimate this vector at node $k$ by suggesting the following cost function as:
    \begin{displaymath}
    J_k(\omegab)=\sum_{l\in \cal N_\mathrm{k}}a_{lk}\mathrm{E}\{||\tilde{d}_{l,i}-b_ld^2_{l,i}-\ub^T_{k,i}\omegab||^2_2\}=
    \end{displaymath}
    \begin{equation}
    \bar{J}_k(\omegab,\bb_k,\db_k),
    \end{equation}
where $\db_k=[d_{l,i}]_{l\in \cal N_\mathrm{k}}$ is the linear ground truth measurement vector of node $k$ with incorporating the measurement noise and is estimated in the second step of the algorithm, and $\bb_k=[b_{l,k}]_{l\in \cal N_\mathrm{k}}$ is the nonlinear coefficients estimated by node $k$. To find $\bb_k$, we use a steepest descent of $\bar{J}_k(\omegab,\bb_k,\hat{\db_k})$ by assuming $\hat{\db_k}$ is known. So, we have
    \begin{equation}
    \hat{\bb}_{k,i}=\hat{\bb}_{k,i-1}-\mu_b\nabla_{\bb_k}\bar{J}=\hat{\bb}_{k,i-1}+\mu_b\cb_k\tilde{\eb}_{k,i}\odot\db^2_{k,i},
    \end{equation}
    where $\cb_k=[c_{lk}]_{l\in \cal N_\mathrm{k}}$, $\tilde{\eb}_{k,i}=[\tilde{e}_{l,i}]_{l\in \cal N_\mathrm{k}}$, and $\db^2=\db\odot\db$.
\item True measurement estimation: In this step, since we estimate $d_{l,i}$ from $\tilde{d}_{l,i}$, i.e., the second-order nonlinear equation, we can equivalently call this step compensation of nonlinear measurements. We use the second-order equation $b_ld^2_{l,i}+d_{l,i}-\tilde{d}_{l,i}=0$. So, neglecting the other incorrect solution, the correct solution is
    \begin{equation}
    \hat{d}_{l,i}=\frac{-1+\sqrt{1+4\hat{b}_l\tilde{d}_{l,i}}}{2\hat{b}_l}.
    \end{equation}
\item Adaptation step: This step is equivalent to a classical adaptation step in DLMS algorithm, i.e.,
    \begin{displaymath}
    \varphib_{k,i}=\omegab_{k,i-1}-\mu\nabla_{\omegab}\bar{J}(\omegab_{k,i-1})=
    \end{displaymath}
    \begin{equation}
    \omegab_{k,i-1}+\mu\sum_{l\in \cal N_\mathrm{k}}c_{lk}\tilde{e}_{l,i}\ub_{l,i},
    \end{equation}
    where $\tilde{e}_{l,i}=\tilde{d}_{l,i}-\hat{b}_l\hat{d}^2_{l,i}-\ub^T_{l,i}\omegab_{k,i-1}$.
\item Compensation of nonlinearity in the intermediate estimation: After exchanging the intermediate estimations of $\varphib_{l,i}$, the received intermediate estimations, i.e., $\tilde{\varphib}_{l,i}=f(\varphib_{l,i})$, should be compensated by the relation of second order nonlinear equation. Similar to step 2, we write
    \begin{equation}
    \hat{\varphi}_{l,i,j}=\frac{-1+\sqrt{1+4\hat{b}_l\tilde{\varphi}_{l,i,j}}}{2\hat{b}_l}.
    \end{equation}
\item Combination step: In this step, we update the final estimation of node $k$ as $    \omegab_{k,i}=\sum_{l\in \cal N_\mathrm{k}}a_{lk}\hat{\varphib}_{l,i}$.
\end{enumerate}
Hence, the SONEC-DLMS algorithm is a five-step diffusion algorithm having three extra steps compared to DLMS algorithm. These three steps are, one step for estimation of nonlinear coefficients and, two other steps for compensation of nonlinearity of measurements and intermediate estimations. We call this algorithm fully-distributed SONEC-DLMS algorithm. The computational complexity of the proposed SONEC-DLMS in terms of number of addition, multiplication, and nonlinear operator of square root, in comparison to DLMS algorithm is depicted in Table 1. It is seen that the computational complexity of the proposed algorithm is approximately three times that of the DLMS algorithm.

\begin{table}[!b]
\caption{Computational Complexity per node $k$ and per iteration of algorithms ($N_k=\mathrm{Card}\{\cal N_\mathrm{k}\}$)}
\label{Table_1}

\centering

 \begin{tabular}{p{14mm}||p{18mm}|p{18mm}|p{10mm}|p{5mm}} \hline
 \Tstrut Algorithm & Add  & Multiplication  & Nonlinear\\\hline \hline

 \Tstrut DLMS
 & $\!\!\! \begin{aligned} L(3N_k-1)& \end{aligned} $	
 & $\!\!\!\begin{aligned} L(3N_k+1)& \end{aligned}$	
 & $\!\!\!\begin{aligned} 0& \end{aligned}$	\\ \hline

 \Tstrut SONEC-DLMS
 &	$\!\!\!\begin{aligned} L(9N_k+1)& \end{aligned}$	
 & $\!\!\!\begin{aligned} L(9N_k+2)& \end{aligned}$
 & $\!\!\!\begin{aligned} 3N_k& \end{aligned}$ \\ \hline

%
%
%

\end{tabular}
 \begin{tabular} {l}
\\
Nonlinear: Nonlinear operators such as square root.
\end{tabular}

\end{table}

To further improve the performance of the SONEC-DLMS algorithm, we propose a semi-distributed SONEC-DLMS algorithm, in which the nonlinear coefficients are estimated in a centralized manner by a fusion center by using training data. This version of the proposed algorithm, which can be considered as a combination of distributed and centralized algorithms, contains only four steps and the nonlinearity estimation step is done separately in a centralized manner. As we will see in the simulation results, the performance of this semi-distributed algorithm is close to that of DLMS without nonlinearity.


\section{Cramer-Rao bound}
\label{sec: CRB}
In this section, the Cramer-Rao bound for estimating the nonlinear coefficient vector $\bb=[b_l]$ and unknown vector $\omegab=\omegab_o$ is derived. Let us define an $(L+N)\times 1$ parameter vector $\thetab=[\omegab^T \bb^T]^T$. We intend to find the CRB of estimated $\thetab$ based on the measurements $\tilde{d}_{l,i}$. All the measurements are defined in a matrix $\Xb=[\xb_1,\xb_2,...,\xb_N]$ where $\xb_l=[\tilde{d}_{l,1},\tilde{d}_{l,2},...,\tilde{d}_{l,I}]^T$ is the total observations of node $l$. From (\ref{eq: dtilde}), we have $\tilde{d}_{l,i}\sim \mathrm{N}(\ub^T_{l,i}\omegab+b_l(\ub^T_{l,i}\omegab)^2,\sigma^2_{\theta,l})$, where $\mathrm{N}(a,b)$ represents the Gaussian distribution with mean $a$ and variance $b$. The Fisher Information Matrix (FIM) of measurements $\Xb$ for estimating $\thetab$ is calculated as $\mathrm{FIM}_{\thetab}=\left(
                         \begin{array}{cc}
                           \Fb_{\omega} & \Fb^T_{b\omega} \\
                           \Fb_{b\omega} & \Fb_b \\
                         \end{array}
                       \right)
$, where $\Fb_{\omega,i,j}=-\mathrm{E}\Big\{\frac{\partial^2\ln p(\Xb;\thetab)}{\partial \omega_i\omega_j}\Big\}$, $\Fb_{b,i,j}=-\mathrm{E}\Big\{\frac{\partial^2\ln p(\Xb;\thetab)}{\partial b_ib_j}\Big\}$, and $\Fb_{b\omega,i,j}=-\mathrm{E}\Big\{\frac{\partial^2\ln p(\Xb;\thetab)}{\partial b_i\omega_j}\Big\}$.

To further proceed and to calculate the likelihood $p(\Xb;\thetab)=p(\xb_1,\xb_2,...,\xb_N;\thetab)$, we write the observation vector $\xb_k$ as
\begin{equation}
\xb_k=\Ub_k\omegab+b_k(\Ub_k\omegab)^2+\zb_k,
\end{equation}
where $\Ub^T_k=[\ub_{k,1}|\ub_{k,2}| ... |\ub_{k,I}]$ and $\zb_k=[\theta_{k,1},\theta_{k,2},...,\theta_{k,I}]^T$. Assuming the independence of measurement noises in different nodes $\zb_k$, we have $\xb_k\sim \mathrm{N}(\Ub_k\omegab+b_k(\Ub_k\omegab)^2,\mathrm{diag}(\sigma^2_{\theta,k}))$. Hence, the log-likelihood can be written as
\begin{displaymath}
\ln p(\Xb;\thetab)=\sum_{k=1}^N \ln p(\xb_k;\thetab)=
\end{displaymath}
\begin{equation}
\sum_{k=1}^N -\frac{N}{2}\ln p(2\pi\sigma^2_{\theta,k})-\frac{1}{2\sigma^2_{\theta,k}}||\xb_k-\Ub_k\omegab+b_k(\Ub_k\omegab)^2||^2.
\end{equation}
So, the partial derivative is $\frac{\partial \ln p(\Xb;\thetab)}{\partial \omega_i\partial\omega_j}=\sum_{k=1}^N\frac{-1}{2\sigma^2_{\theta,k}}\frac{\partial}{\partial\omega_i\omega_j}||\xb_k-\Ub_k\omegab-b_k(\Ub_k\omegab)^2||^2
$. To further proceed, we define $\rb_k=\Ub_k\omegab+b_k(\Ub_k\omegab)^2$.

Let $A=\frac{\partial^2}{\partial\omega_i\omega_j}(\rb^T_k\rb_k)=\frac{\partial^2}{\partial\omega_i\omega_j}\sum_{p=1}^I r^2_{k,p}$, then some calculations lead to
\begin{equation}
\label{eq: pp}
\frac{\partial^2}{\partial\omega_i\omega_j}(\rb^T_k\rb_k)=2\rb^T_k\frac{\partial^2 \rb_k}{\partial\omega_i\omega_j}+2(\frac{\partial \rb_k}{\partial\omega_i})^T(\frac{\partial \rb_k}{\partial\omega_j}).
\end{equation}
Using (\ref{eq: pp}) and taking expectations and considering that $\mathrm{E}(\xb_k)=\rb_k$, we have
\begin{equation}
\Fb_{\omega,i,j}=\sum_{k=1}^N\frac{1}{2\sigma^2_{\theta,k}}(\frac{\partial \rb_k}{\partial\omega_i})^T(\frac{\partial \rb_k}{\partial\omega_j})=\sum_{k=1}^N\frac{1}{2\sigma^2_{\theta,k}}\pb^T_{k,i}\pb_{k,j},
\end{equation}
where $\pb_{k,i}=\frac{\partial \rb_k}{\partial\omega_i}$. Some calculations show that we have
\begin{equation}
\label{eq: pki}
\pb_{k,i}=\ub_{k,:,i}\odot[1+2b_k(\Ub_k\omegab)],
\end{equation}
where $\ub_{k,:,i}=[u_{k,1,i}, u_{k,2,i},...,u_{k,I,i}]^T$. Similar calculations can lead to


\begin{equation}
\label{eq: Fw}
\Fb_{\omega,i,j}=\sum_{k=1}^N\frac{1}{\sigma^2_{\theta,k}}\pb^T_{k,i}\pb_{k,j}, \Fb_{b,i,j}=\sum_{k=1}^N\frac{1}{\sigma^2_{\theta,k}}\acute{\pb}^T_{k,i}\acute{\pb}_{k,j},
\end{equation}
\begin{equation}
\label{eq: Fbw}
\Fb_{b\omega,i,j}=\sum_{k=1}^N\frac{1}{\sigma^2_{\theta,k}}\acute{\pb}^T_{k,i}{\pb}_{k,j},
\end{equation}
where
\begin{equation}
\label{eq: pprimki}
\acute{\pb}_{k,i}=\Bigg\{\begin{array}{ll}
0 \quad\quad\quad k\neq i,\\
(\Ub_k\omegab)^2 \quad k=i.
                  \end{array}
\end{equation}

\section{Simulation Results}
\label{sec: Simulation}
In this section, some experiments are performed to investigate the performance of the proposed SONEC-DLMS algorithm in a distributed network with nonlinear sensors. The simulation setup is as follows. The WSN consists of $N=16$ sensors. The network is selected the same as the one introduced in \cite{Zayy22CSSP}. The sensors of WSN are collected a second order nonlinear measurement of a $L\times 1$ unknown vector $\omegab^o$ with L=20. The unknown vector elements and the elements of the regression vectors are chosen from normal distribution. For the nonlinear coefficients, we use a uniform random variable $b_l\sim U(-b_{l,\mathrm{max}},0)$. Unless otherwise stated, we use $b_{l,\mathrm{max}}=0.4$ in the simulations. The proposed distributed estimation algorithm aims to estimate the unknown vector in an adaptive manner. For the background noise $n_{k,i}$, we use a zero-mean Gaussian distribution with standard deviation equal to 0.045. The performance metric for evaluating the performance of the proposed algorithm is the mean square deviation (MSD) criterion defined as $\mathrm{MSD}(\mathrm{dB})=20~\mathrm{log}_{10}(||\omegab-\omegab_o||_2)$. The number of Monte Carlo simulations is 100 independent runs and the results are averaged over all runs. The combination coefficients $a_{lk}$ and $c_{lk}$ are chosen by the uniform policy \cite{Sayed14}.

The performance of the proposed SONEC-DLMS algorithm with its two different versions which are fully-distributed and semi-distributed is investigated. Furthermore, we have included a version of the proposed algorithm that utilizes compensation solely in the combination step. However, we have excluded another variant of the algorithm that applies compensation in the adaptation step but not in the combination step, as it failed to converge in the simulations. The step-sizes $\mu_k$ and $\mu_b$ have been carefully selected to minimize the final mean square deviation (MSD), with values of $\mu_k = 0.01$ and $\mu_b = 0.005$. The MSD of estimating the unknown vector $\omegab_o$ versus iteration number is depicted in Fig. 1. As it can be seen, the fully distributed SONEC-DLMS algorithm performs better than DLMS by at least 7dB. It shows that the performance of the semi-distributed SONEC-DLMS algorithm is close to that of the DLMS without nonlinearity. It demonstrates that there is a gap of at least 10dB between the CRB and the SONEC-DLMS. It also shows that the upper-bound is not tight, but is less than $-10dB$. Moreover, the MSD of estimating the nonlinear coefficient vector $\bb$ versus iteration number is shown in Fig. 2. It shows a large gap between the performance of SONEC-DLMS algorithms and CRB.

\begin{figure}[tb]
\begin{center}
\includegraphics[width=7cm]{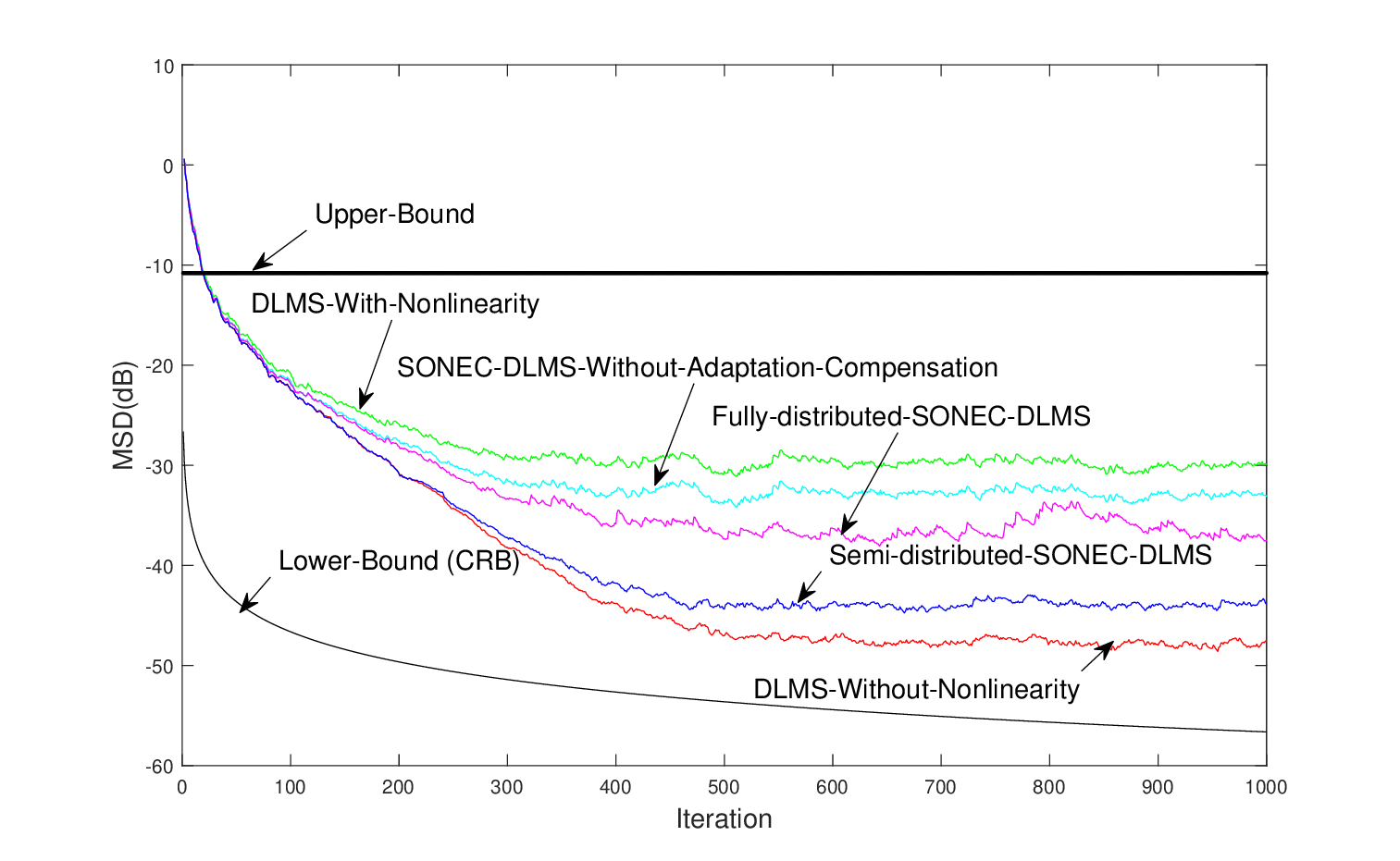}
\end{center}
\caption{Variation of MSD for estimating the unknown vector $\omegab_o$ for DLMS with nonlinearity, DLMS without nonlinearity, fully-distributed SONEC-DLMS, semi-distributed SONEC-DLMS, and CRB.}
\label{fig1}
\end{figure}

\begin{figure}[tb]
\begin{center}
\includegraphics[width=7cm]{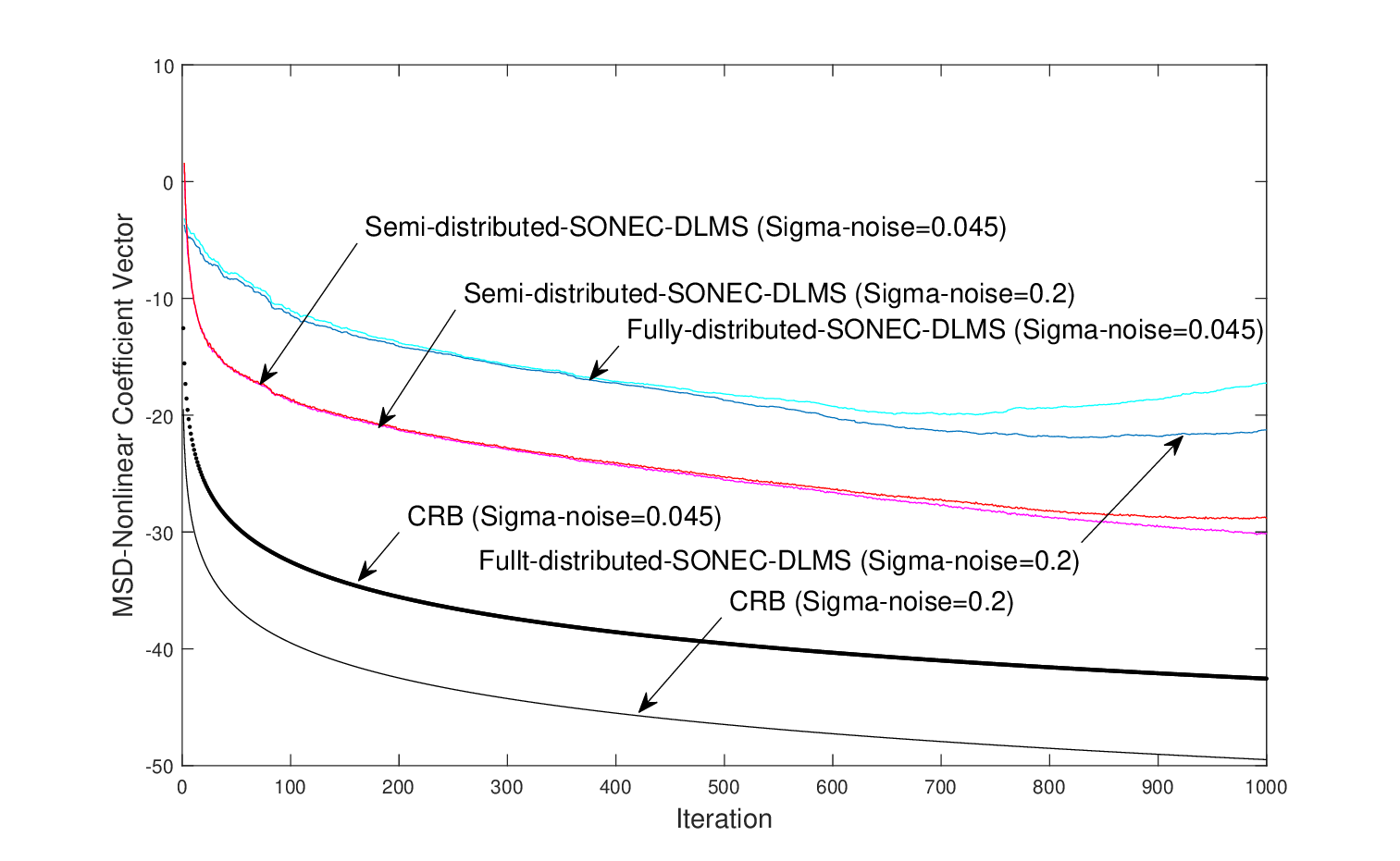}
\end{center}
\caption{Variation of MSD for estimating the nonlinear coefficient vector $\bb$ for fully-distributed SONEC-DLMS, semi-distributed SONEC-DLMS, and CRB.}
\label{fig2}
\end{figure}



%

\section{Conclusion}
\label{sec: con}
In this paper, a solution for improving the performance of distributed estimation in the presence of nonlinearities was provided. The solution is to use a nonlinearity estimation and nonlinearity compensation units. Moreover, an upper bound for the error due to nonlinearity is obtained. Also, the CRB of the problem of distributed estimation in the presence of second-order nonlinearity was calculated. Simulation results show the effectiveness of the proposed algorithm. The future work could be working on the low resolution messages between sensors \cite{Ciu20}.

\section*{Appendix 1}
To calculate $\fb_{l,i}=\mathrm{E}\{\pb_{l,i}\}$, we can write
\begin{displaymath}
\fb_{l,i}=\sum_{l^{'}\in \cal N_\mathrm{l}}c_{l^{'},l}\mathrm{E}\{e_{l^{'},i}\ub_{l^{'},i}\}=\sum_{l^{'}\in \cal N_\mathrm{l}}c_{l^{'},l}\mathrm{E}\{(d_{l^{'},i}-\ub^T_{l^{'},i}\omegab_{l^{'},i-1})\ub_{l^{'},i}\}=
\end{displaymath}
\begin{displaymath}
\sum_{l^{'}\in \cal N_\mathrm{l}}c_{l^{'},l}\mathrm{E}\{\ub^T_{l^{'},i}(\omegab^{o}-\omegab_{l^{'},i-1})\ub_{l^{'},i}+\vb_{l^{'},i}\ub_{l^{'},i}\}=
\end{displaymath}
\begin{equation}
-\sum_{l^{'}\in \cal N_\mathrm{l}}c_{l^{'},l}\mathrm{E}\{\ub^T_{l^{'},i}\tilde{\omegab}_{l^{'},i-1}\ub_{l^{'},i}\}
\end{equation}
To calculate $\sbb=\mathrm{E}\{\ub^T_{l^{'},i}\tilde{\omegab}_{l^{'},i-1}\ub_{l^{'},i}\}$, we nominate $\rb=\ub_{l^{'},i}$ and $\vb=\tilde{\omegab}_{l^{'},i-1}$ for simplicity. So, we have
\begin{equation}
s_k=\mathrm{E}\{\rb^T\vb r_k\}=\mathrm{E}\{(r_1v_1+...+r_Lv_L)r_k\}=\mathrm{E}\{r^2_kv_k\},
\end{equation}
since $r_k=u_{l^{'},i-1,k}$ and $r_j=r_k=u_{l^{'},i-1,j}$ for $k\neq j$ are uncorrelated. Hence, we have
\begin{equation}
\mathrm{E}\{\ub^T_{l^{'},i}\tilde{\omegab}_{l^{'},i-1}\ub_{l^{'},i}\}=-\mathrm{E}\{\rb^2\odot\vb\}=-\sigma^2_u\tilde{\tilde{\omegab}}_{l^{'},i-1}.
\end{equation}
Therefore, we reach to (\ref{eq: fli}).
To calculate $\gb_{l,i}=\mathrm{E}\{\Delta\varphib_{l,i}\}$, we can write
\begin{equation}
\gb_{l,i}=\mathrm{E}\{\Delta\varphib_{l,i}\}=\sum_{l^{'}\in \cal N_\mathrm{l}}c_{l^{'},l}\mathrm{E}\{d^2_{l^{'},i}\ub_{l^{'},i}\}=\sum_{l^{'}\in \cal N_\mathrm{l}}c_{l^{'},l}b_{l^{'}}\qb_{l^{'},i},
\end{equation}
where $\qb_{l^{'},i}\triangleq \mathrm{E}\{d^2_{l^{'},i}\ub_{l^{'},i}\}$. Then, we have
\begin{displaymath}
\qb_{l^{'},i}=\mathrm{E}\{(\ub^T_{l^{'},i}\omegab_{l^{'},i-1}+v_{l^{'},i})^2\ub_{l^{'},i}\}=
\end{displaymath}
\begin{equation}
\label{eq: qq}
\mathrm{E}\{(\ub^T_{l^{'},i}\omegab_{l^{'},i-1})^2\ub_{l^{'},i}\}+2\mathrm{E}\{v_{l^{'},i}(\ub^T_{l^{'},i}\omegab_{l^{'},i-1})\ub_{l^{'},i}\}\}+\mathrm{E}\{v^2_{l^{'},i}\ub_{l^{'},i}\},
\end{equation}
where the second and third term in (\ref{eq: qq}) are zero. So, we have
\begin{equation}
\qb_{l^{'},i}=\mathrm{E}\{(\ub^T_{l^{'},i}\omegab_{l^{'},i-1})^2\ub_{l^{'},i}\}=\mathrm{E}\{(\ub^T\wb)^2\ub\}=\mathrm{E}\{(\wb^T\ub\ub^T\wb)\ub\},
\end{equation}
where we define $\ub=\ub^T_{l^{'},i}$ and $\wb=\omegab_{l^{'},i-1}$ for simplicity. Then, we have
\begin{equation}
q_{l^{'},i,r}=\mathrm{E}\{(\wb^T\ub\ub^T\wb)u_r\}=\mathrm{E}\Big\{\sum_{j=1}^L\sum_{k=1}^L w_ju_jw_ku_ku_r\Big\}=\mathrm{E}\{f_r\},
\end{equation}
where we have
\begin{displaymath}
f_r=\sum_{j=1}^L\sum_{k=1}^L w_ju_jw_ku_ku_r=
\end{displaymath}
\begin{equation}
\sum_{k=1,k\neq r}^L\sum_{j=1,j\neq r}^L w_ju_jw_ku_ku_r+\sum_{j=1,j\neq r}^L w_ju_jw_ru_ru_r+\sum_{k=1}^L w_ju_rw_ru_ku_r.
\end{equation}
Hence, simple calculations show that
\begin{equation}
q_{l^{'},i,r}=\sum_{k=1,k\neq r}^L\sum_{j=1,j\neq r}^L \mathrm{E}\{u_ju_ku_r\}w_jw_k+\sum_{j=1,j\neq r}^L \mathrm{E}\{u_ju^2_r\}w_jw_r+\sum_{k=1}^L \mathrm{E}\{u^2_ru_k\}w_jw_k.
\end{equation}
Using the whiteness of $\ub$, some simple calculations show that $q_{l^{'},i,r}=0$. Hence, we reach to (\ref{eq: gli}).
To calculate $\kb_{l,i}$, we have
\begin{equation}
\label{eq: abov}
\kb_{l,i}=\mathrm{E}\{(\varphib_{l,i}+\Delta\varphib_{l,i})^2\}=\mathrm{E}\{\varphib^2_{l,i}\}+2\mathrm{E}\{\varphib_{l,i}\odot\Delta\varphib_{l,i}\}+\mathrm{E}\{(\Delta\varphib_{l,i})^2\}.
\end{equation}
The second term in (\ref{eq: abov}) is zero if we assume the uncorelatedness of $\varphib_{l,i}$ and $\Delta\varphib_{l,i}$ and also since we proved that $\mathrm{E}\{\Delta\varphib_{l,i}\}=\mathbf{0}$ in the current appendix. So, we have
\begin{equation}
\kb_{l,i}=\mathrm{E}\{\varphib^2_{l,i}\}+\mathrm{E}\{(\Delta\varphib_{l,i})^2\}=\hb_{l,i}+\rb_{l,i}.
\end{equation}

\section*{Appendix 2}
To compute $\hb_{l,i}=\mathrm{E}\{\varphib^2_{l,i}\}$, we have
\begin{equation}
\label{eq: pb2}
\hb_{l,i}=\omegab^2_{l,i-1}+2\mu\omegab_{l,i-1}\odot\mathrm{E}\{\pb_{l,i}\}+\mu^2\mathrm{E}\{\pb^2_{l,i}\}=\omegab^2_{l,i-1}+2\mu\omegab_{l,i-1}\odot\fb_{l,i}+\mu^2\mathrm{E}\{\pb^2_{l,i}\}.
\end{equation}
To calculate $\mathrm{E}\{\pb^2_{l,i}\}$ in (\ref{eq: pb2}), some calculations show that
\begin{equation}
\pb^2_{l,i}=\sum_{l^{'}\in \cal N_\mathrm{l}}c_{l^{'},l}\ub^T_{l^{'},i}\tilde{\omegab}_{l^{'},i-1}\ub_{l^{'},i}\odot\sum_{l^{''}\in \cal N_\mathrm{l}}c_{l^{''},l}\ub^T_{l^{''},i}\tilde{\omegab}_{l^{''},i-1}\ub_{l^{''},i}.
\end{equation}
Now, $t_{l,i,r}=\mathrm{E}\{p^2_{l,i,r}\}$ is equal to
\begin{displaymath}
t_{l,i,r}=\mathrm{E}\{\Big(\ub^T_{l^{'},i}\tilde{\omegab}_{l^{'},i-1}\ub_{l^{'},i,r}\Big)\Big(\ub^T_{l^{''},i}\tilde{\omegab}_{l^{''},i-1}\ub_{l^{''},i,r}\Big)\}=
\end{displaymath}
\begin{equation}
\sum_{l^{'}\in \cal N_\mathrm{l}}\sum_{l^{''}\in \cal N_\mathrm{l}}c_{l^{'},l}c_{l^{''},l}\mathrm{E}\Big\{(\ub^T_{l^{'},i}\tilde{\omegab}_{l^{'},i-1})(\ub^T_{l^{''},i}\tilde{\omegab}_{l^{''},i-1})u_{l^{'},i,r}u_{l^{''},i,r}\Big\}.
\end{equation}
Therefore, from (\ref{eq: pb2}), we have
\begin{equation}
\hb_{l,i}=\omegab^2_{l,i-1}+2\mu\omegab_{l,i-1}\odot\fb_{l,i}+\mu^2\tb_{l,i}.
\end{equation}

To calculate $\rb_{l,i}=\mathrm{E}\{\Delta^2\varphib_{l,i}\}$, we can write
\begin{equation}
\Delta^2\varphib_{l,i,r}=\sum_{l^{'}\in \cal N_\mathrm{l}}\sum_{l^{''}\in \cal N_\mathrm{l}}c_{l^{'},l}c_{l^{''},l}b_{l^{'}}b_{l^{''}}d^2_{l^{'},i}d^2_{l^{''},i}u_{l^{'},i,r}u_{l^{''},i,r}.
\end{equation}
So, we have
\begin{displaymath}
r_{l,i,r}=\sum_{l^{'}\in \cal N_\mathrm{l}}\sum_{l^{''}\in \cal N_\mathrm{l}}c_{l^{'},l}c_{l^{''},l}b_{l^{'}}b_{l^{''}}\mathrm{E}\{d^2_{l^{'},i}d^2_{l^{''},i}u_{l^{'},i,r}u_{l^{''},i,r}\}=
\end{displaymath}
\begin{equation}
\label{eq: rlir}
\sum_{l^{'}\in \cal N_\mathrm{l}}\sum_{l^{''}\in \cal N_\mathrm{l}}c_{l^{'},l}c_{l^{''},l}b_{l^{'}}b_{l^{''}}\mathrm{E}\{(\ub^T_{l^{'},i}\omegab_{l^{'},i-1}+v_{l^{'},i})^2(\ub^T_{l^{''},i}\omegab_{l^{''},i-1}+v_{l^{''},i})^2u_{l^{'},i,r}u_{l^{''},i,r}\}.
\end{equation}
To calculate
\begin{equation}
\mathrm{A}_{l^{'},l^{''},r}=\mathrm{E}\{(\ub^T_{l^{'},i}\omegab_{l^{'},i-1}+v_{l^{'},i})^2(\ub^T_{l^{''},i}\omegab_{l^{''},i-1}+v_{l^{''},i})^2u_{l^{'},i,r}u_{l^{''},i,r}\},
\end{equation}
for simplicity, it is re-written as
\begin{displaymath}
\mathrm{A}_{l^{'},l^{''},r}=\mathrm{E}\{(\ub^T_{l^{'}}\wb_{l^{'}}+v_{l^{'}})^2(\ub^T_{l^{''}}\wb_{l^{''}}+v_{l^{''}})^2 u_{l^{'},r}u_{l^{''},r}\}=
\end{displaymath}
\begin{equation}
\mathrm{E}\Big\{\Big[(\ub^T_{l^{'}}\wb_{l^{'}})^2+2\vb_{l^{'}}\ub^T_{l^{'}}\wb_{l^{'}}+v^2_{l^{'}} \Big]\Big[(\ub^T_{l^{''}}\wb_{l^{''}})^2+2\vb_{l^{''}}\ub^T_{l^{''}}\wb_{l^{''}}+v^2_{l^{''}} \Big]u_{l^{'},r}u_{l^{''},r}\Big\}.
\end{equation}
Then, some calculations lead to
\begin{displaymath}
\mathrm{A}_{l^{'},l^{''},r}=\mathrm{E}\Big\{\Big[(\ub^T_{l^{'}}\wb_{l^{'}})^2(\ub^T_{l^{''}}\wb_{l^{''}})^2u_{l^{'},r}u_{l^{''},r}\Big]\Big\}+\mathrm{E}\Big\{\Big[(\ub^T_{l^{'}}\wb_{l^{'}})^2v^2_{l^{''}}\Big]u_{l^{'},r}u_{l^{''},r}\Big\}+
\end{displaymath}
\begin{displaymath}
4\mathrm{E}\Big\{\Big[v_{l^{'}}v_{l^{''}}(\ub^T_{l^{'}}\wb_{l^{'}})(\ub^T_{l^{''}}\wb_{l^{''}}) \Big]u_{l^{'},r}u_{l^{''},r} \Big\}+\mathrm{E}\Big\{\Big[v^2_{l^{'}}(\ub^T_{l^{''}}\wb_{l^{''}})^2\Big]u_{l^{'},r}u_{l^{''},r}\Big\}+
\end{displaymath}
\begin{equation}
\mathrm{E}\{v^2_{l^{'}}v^2_{l^{''}}u_{l^{'},r}u_{l^{''},r}\}.
\end{equation}
It is easy to show that if $l^{''}\neq l^{'}$, we have $\mathrm{A}_{l^{'},l^{''},r}=\mathbf{0}$. Also, for $l^{''}=l^{'}$, we have
\begin{displaymath}
\mathrm{A}_{l^{'},l^{'},r}=\mathrm{E}\{(\ub^T_{l^{'}}\wb_{l^{'}})^4u^2_{l^{'},r}\}+\mathrm{E}\{(\ub^T_{l^{'}}\wb_{l^{'}})^2v^2_{l^{'}}u^2_{l^{'},r}\}+
\end{displaymath}
\begin{equation}
5\sigma^2_v\mathrm{E}\{(\ub^T_{l^{'}}\wb_{l^{'}})^2u^2_{l^{'},r}\}+\sigma^2_v+3\sigma^4_v\sigma^2_u.
\end{equation}
Finally, from (\ref{eq: rlir}), we have
\begin{equation}
r_{l,i,r}=\sum_{l^{'}\in \cal N_\mathrm{l}}c^2_{l^{'},l}b^2_{l^{'}}\mathrm{A}_{l^{'},l^{'},r}.
\end{equation}

\end{document}